\documentclass[a4paper,aps, prl,amsmath,amssymb, superscriptaddress, showpacs, twocolumn, floatfix]{revtex4-1}
\usepackage{graphicx, braket, mathrsfs, nicefrac, epsfig, multirow, ulem, verbatim, subfigure}

\begin{document}
\title{ Quantum synchronization and transresistance quantization in superconducting devices}

\author{A.~M.~Hriscu}
\affiliation{Kavli Institute of Nanoscience, Delft University of
Technology,\\
PO Box 5046, 2600 GA Delft, The Netherlands}
\author{Yu.~V.~Nazarov}
\affiliation{Kavli Institute of Nanoscience, Delft University of
Technology,\\
PO Box 5046, 2600 GA Delft, The Netherlands}

\begin{abstract}
We show theoretically the possibility of quantum synchronization of Josephson and Bloch oscillations in a superconducting device. One needs an $LC$ oscillator to achieve exponentially small rate of synchronization errors. The synchronization leads to quantization of transresistance similar to that in (Fractional) Quantum Hall Effect.  
\end{abstract}

\maketitle


One of the most interesting discoveries of XX century was the perfect (fractional) quantization of Hall transresistance in rather imperfect 2DEG semiconducting samples\cite{Klitzing}. The resistance as a function of electron density and magnetic field tends to be close to plateaus with values
\begin{equation}
\label{eq:QHE}
R = \frac{V}{I} = \frac{2\pi \hbar}{e^2} \frac{m}{n}
\end{equation}
$n$, $m$ being integer numbers. The accuracy is so good as to enable numerous metrological applications \cite{QHEMetrology1, QHEMetrology2}. The physical explanation of the effect is the commensurability of electron density and density of the magnetic flux penetrating the sample, this taking place any time the ratio of  numbers of elementary charges and 
flux quanta in the structure is a rational fraction $n/m$. 

Quantum Hall samples are macroscopic involving infinitely many degrees of freedom. Shortly after the discovery, Likharev and Zorin \cite{LZ} hypothesized that similar resistance quantization may occur in a Josephson-junction superconducting device encompassing only few quantum degrees of freedom. They foresaw it as a result of quantum {\it synchronization} of Bloch \cite{Bloch1928} and Josephson \cite{Josephson1962} oscillations in two junctions. The Josephson frequency $\omega_J = 2e V_{\cal O}/\hbar$ is proportional to the average voltage dropping at one of the junctions while the Bloch frequency $\omega_B = \pi I_{\cal O}/e$ is proportional to the average current in another junction. A synchronization condition of the two oscillations, $n \omega_J = m\omega_B$ results in 
\begin{equation}
R = \frac{V_{\cal O}}{I_{\cal O}} = \frac{\pi \hbar}{2 e^2} \frac{m}{n}.
\end{equation}
The resistance quantum is modified in comparison with Eq. \ref{eq:QHE} manifesting the double charge $2e$ of Cooper pairs in superconductors. Unfortunately, the original device suggestion \cite{LZ} does not work. The reason of the failure seems fundamental. The quantities to be synchronized, the charge and flux in the device are canonically conjugated variables. Quantum mechanics forbids them to be simultaneously certain, and the synchronization is expected to be destroyed by quantum fluctuations.

A recent outburst of theoretical and experimental activities concerns quantum-coherent phase slips in thin nanowires \cite{Review}. On theoretical side, a concept of phase-slip (PS) junction has emerged \cite{MooijHarmans,MooijNazarov}. Such junction is exactly dual to a common Josephson junction with respect to charge-flux conjugation. This inspired the proposals of novel superconducting devices \cite{HriscuPRL, HriscuPRB, Vanevic2012}. Very recently, a PS qubit on InO nanowires has been realized \cite{Astafiev}. Relevant experimental developments include observation of the predicted phenomena: phase-slips in Josephson junction chains \cite{Guichard2010,Manucharyan2012}, Bloch oscillations \cite{Arutyunov2012}, and charge sensitivity \cite{Zorin2012}.

In this Letter, we demonstrate that combining PS and Josephson junctions in a single device solves the problem of quantum synchronization. A necessary element of the device appears to be an $LC$ oscillator with high quality factor $Q$. With this, one can make the rate $\Gamma$ of synchronization errors exponentially small, 
$-\ln\Gamma \simeq Q$ thereby achieving exponential accuracy of the resistance quantization. Importantly, the device suggested can be also used as both voltage and current standard, thereby closing the metrological triangle \cite{Triangle}.

\begin{figure} [!h]
\includegraphics[width=0.7\columnwidth]{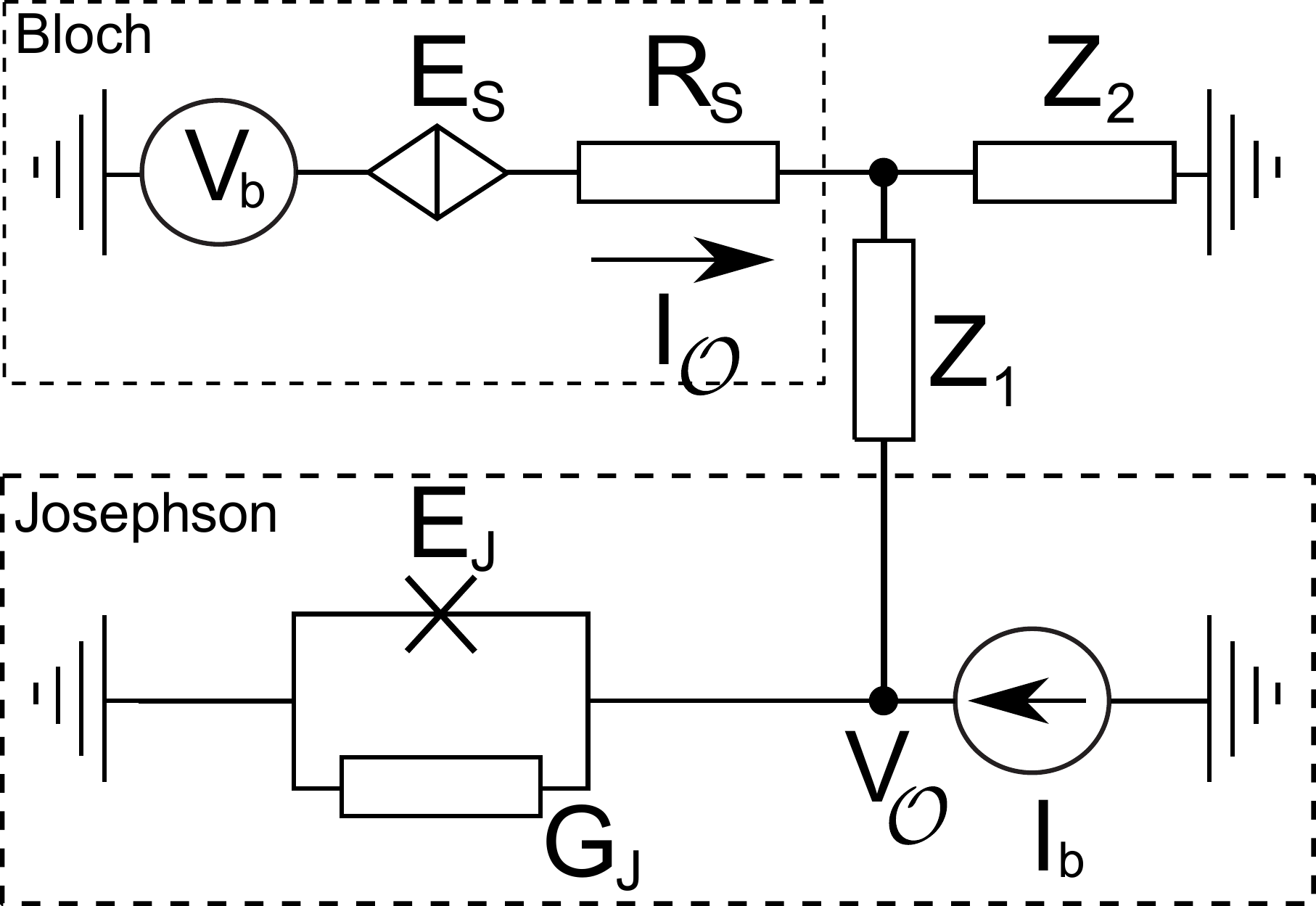}
\caption{ A PS (marked with diamond) and Josephson (cross) junction embedded into a general linear circuit. The circuit parts in dashed boxes generate Bloch and Josephson oscillations while the (frequency-dependent) resistors $Z_1$ and $Z_2$ provide the coupling between the parts. The circuit is controlled with voltage and current sources $V_b, I_b$. The d. c. output voltage and current $V_{\cal O},I_{\cal O}$ manifest the quantized transresistance $R = V_{\cal O}/I_{\cal O}$.}
\label{Fig-General}
\end{figure}

To appreciate the difficulty of quantum synchronization, we consider first a PS and a Josephson junction embedded in a general linear circuit that can be represented with four (frequency-dependent) resistors(Fig. \ref{Fig-General}). The circuit parts in the dashed boxes represent the Bloch and Josephson oscillators. Let us first consider them separately by setting two coupling resistors $Z_{1,2}$ to $Z_1 =\infty$, $Z_2=0$. The Josephson  part is then a common \cite{LikharevBook} Josephson current-biased junction shunted by the conductor $G_J$. 
If the bias current exceeds the critical one, 
$I_b > I_C \equiv{2eE_J}/{\hbar}$, the circuit produces voltage oscillations with frequency $\omega_J = \frac{2e V_{\cal O}}{\hbar} = \frac{2e }{\hbar G_J}\sqrt{I^2_b -I_C^2}$, $V_{\cal O}$ being the time-averaged voltage across the junction. The energy accumulated in the oscillation is of the order of Josephson energy $E_J$. To have a well-defined semiclassical oscillation, we shall require that the energy accumulated by far exceeds the quantum frequency scale $\hbar \omega_J$. The latter can be regarded as an effective noise temperature $T^*_J$ characterizing the quantum fluctuations in the circuit (we neglect the thermal fluctuations assuming sufficiently small temperature). The condition $E_J \gg T^*_{J}$ amounts to $G_J \gg e^2/\hbar$, the conductance must be high at quantum scale.

The Bloch oscillator is understood with using the duality transformation between the phase and charge \cite{MooijNazarov}. Upon such a transformation, the Josephson junction is replaced by a PS junction, the current bias by the voltage bias, and the parallel conductor becomes a series resistor $R_S$. 
Bloch oscillations  occur provided the bias voltage exceeds the critical voltage of the junction, $V_b>V_C=\pi E_S/e$. Their frequency  $\omega_B = \frac{\pi I_{\cal O}}{e} = \frac{\pi }{e R_S} \sqrt{V_b^2 -V_C^2}$, is related to  $I_{\cal O}$, the average current in the junction. To have a well-defined semiclassical oscillation, we shall require that the energy accumulated $\simeq E_S$ by far exceeds the effective noise temperature $T^*_B \simeq \hbar \omega_B$. This gives $R_S \gg \hbar/e^2$: for a PS junction, it is the resistance that must be high at quantum scale.

Let us now couple the circuits. The main effect of the coupling is the transfer of oscillating voltage/current from Josephson/Bloch to Bloch/Josephson part, whereby the voltage/current is multiplied with the amplification coefficient $K(\omega) \equiv Z_2/(Z_1+Z_2)$.
Besides, the effective resistance/conductance of Bloch/Josephson part is modified, $\delta R_S = Z_2Z_1/(Z_2+Z_1)$, $\delta G_J = 1/(Z_2+Z_1)$. In order to preserve well-defined oscillations, we require this modification to be small, $\delta R_S \ll R_S, \delta G_J \ll G_J$. 

We estimate the energy scale $E_{sp}$ associated with the coupling and synchronization of the oscillations as a product of oscillating voltage and current in each device times oscillation period, assuming $\omega_B \simeq \omega_J \simeq \omega$,
$E_{sp} \simeq \tilde{I}_{\cal O} K(\omega) \tilde{V}_{\cal O}/\omega$. It is important to recall that the oscillating quantities are fundamentally related to frequency, $\tilde{I}_{\cal O} \simeq e \omega$, $ \tilde{V}_{\cal O} \simeq \hbar omega/e $. With this, 
$E_{cp} \simeq K \hbar \omega$. A generic estimation for $K$ is $K \lesssim 1$. Indeed, for real impedances $Z_{1,2}$ $K<1$. In this case $E_{cp} \lesssim T^*_{B,J}$ and the envisaged synchronization in a {\it general} circuit is destroyed by quantum fluctuations. 
 
To overcome this, we need large $K$. An active amplifying circuit could provide this but brings extra noise that increases the fluctuations. The main idea of this Letter is to use a passive amplifying circuit, an LC-oscillator, replacing $Z_1$ with a capacitor $C$
and $Z_2$ with an inductor $L$ (Fig. 2). With this, $K(\omega)\gg 1$ near the resonant frequency $\Omega \equiv (LC)^{1/2}$.   
Assuming that a small real part of $Z_2$ gives rise to a finite quality factor $Q$ of the oscillator, $K=  (2(\omega/\Omega-1) +i Q)^{-1}$ at $\omega \approx \Omega$. The maximum value of $K$ is thus limited by $Q$, leading to $E_{cp} \simeq Q \hbar \omega \gg T^*_{J,B}$. We expect the synchronization errors to be related to activation over this energy barrier and thus to occur at exponentially small rate $\simeq \exp(-E_{cp}/T^*)\simeq \exp(-\alpha Q)$, $\alpha$ being a coefficient of the order of $1$. We stress and prove further that the synchronization takes place in a rather broad interval of frequencies near $\Omega$: the Josephson and Bloch oscillations are thus synchronized with each other rather than with the LC oscillations.

\begin{figure} [!h]
\includegraphics[width=0.5\columnwidth]{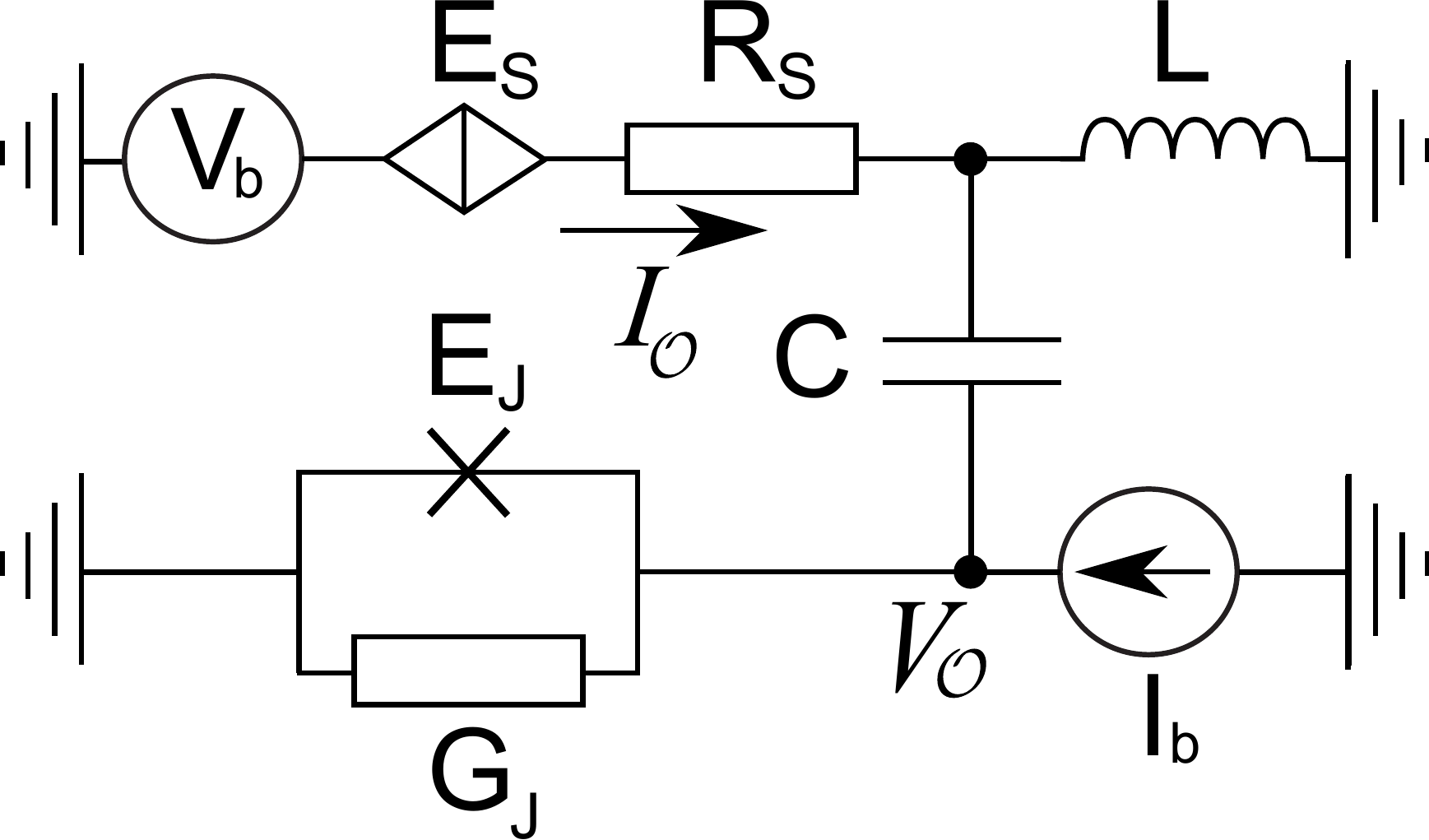}
\caption{Coupling the Josephson and Bloch parts with an $LC$ oscillator results in $K \gg 1$ close to the resonant frequency $\Omega$ and thus enables quantum synchronization.}
\label{Fig-PS}
\end{figure}

The effective quality factor in our circuit is in fact limited by dissipation in $R_S,G_J$. The conditions of non-obtrusive coupling $\delta G_J \ll G_J, \delta R_S \ll R_S$ imply that $Q \ll {\rm min}(G_J z_0, R_S/z_0)$. In fact, the corresponding equality estimates the maximum effective quality factor $Q^{-1}_{m} = 1/G_J z_0 + z_0/R_S$.
A simple optimization of $Q_m$ is to chose the oscillator impedance $z_0 = \sqrt{R_S/G_J}$, so that $Q_m =\sqrt{R_S G_J}/2$. 

Synchronization persists in a finite interval of frequencies $\omega_B(V_b)$, $\omega_J(I_b)$ near the line where those satisfy a given fractional ratio $\omega_B/\omega_J=n/m$. To estimate the width of the interval, we compare $E_{cp}$ with an energy scale characterizing the frequency deviation, which is either $(\Delta \omega_B/\omega_B) E_S$ or $(\Delta \omega_J/\omega_J) E_S$, this leads to $(\Delta \omega_B/\omega_B) \simeq K/(R_S e^2/\hbar)$, $(\Delta \omega_J/\omega_J) \simeq K/(G_J \hbar/e^2)$ We see that for the limiting $Q$ and at frequencies close to $\Omega$ the width of these intervals may become comparable with the frequency itself, $\Delta \omega_J/\omega_J, \Delta \omega_B/\omega_B \simeq 1$. 

In the remainder of the Letter, we support these qualitative estimations with quantitative illustrations. 

The adequate quantum description of the circuit involves two variables: superconducting phase drop at the Josephson junction $\hat \phi$ and dimensionless charge $\hat q=\frac{\pi}{e}\hat {\cal Q}$ flown in the PS junction. It is obtained in the framework of Keldysh action formalism \cite{Schoen} where variables are doubled  $\hat \phi \to \phi^{\pm}(t)$, $\hat q \to q^{\pm}(t)$ corresponding to two parts of the Keldysh contour. It is convenient to use "classical" and "quantum" variables defined as $2\phi,\phi_d=(\phi^+ \pm\phi^-)$, $2q,q_d=(q^+ \pm q^-)$. The total Keldysh action 
$$
{\cal S} = {\cal S}_B + {\cal S}_J + {\cal S}_{cp} + {\cal S}_N  
$$
is contributed by the Bloch and Josephson parts, 
\begin{align}
{\cal S}_J &= \int dt \left( 2 E_J \sin\phi \sin\frac{\phi_d}{2} -\frac{I_b}{2e} \phi_d  +  \dot{\phi} \phi_d\frac{G_J}{4e^2}\right) \\
{\cal S}_B &= \int dt \left( 2 E_S \sin q \sin\frac{q_d}{2} -\frac{e V_b}{\pi}  q_d  + \dot{q} q_d \frac{e^2 R_S}{\pi^2}\right) \end{align}
the coupling part
\begin{align}
{\cal S}_{cp} =\int \frac{d\omega}{2\pi} 
\left(\phi^d_{-\omega} \frac{\delta G}{4e^2}\left(\dot{\phi}\right)_\omega + q^d_{-\omega} \frac{e^2 \delta R}{\pi^2}\left(\dot{q}\right)_\omega \right.+ \\
\left. \frac{K(\omega)}{2\pi}\left(q^d_{-\omega}\left(\dot{\phi}\right)_\omega - \phi^d_{-\omega}\left(\dot{q}\right)_\omega\right)\right) \nonumber
\end{align}  
and the noise part that is quadratic in $q_d,\phi_d$ and satisfies fluctuation-dissipation theorem (see \cite{supp} for concrete expressions). The resulting action is non-local in time and therefore cannot be treated exactly.

To start with, we study the resulting saddle-point classical equations \cite{supp} neglecting the noise. This approximation gives a good estimation of the positions and widths of the synchronization domains while disregarding rounding of large and vanishing of small domains. Typical results are presented in Fig. 3. For this plot, we made (mostly for esthetic reasons) a symmetric choice of parameters 
$E_S=E_J$, $G_J \hbar \pi /4 e^2 = e^2 R_S/\pi \hbar$,
so that output current and voltage, and correspondingly the oscillation frequencies are symmetric in
 the plane of $V_b$ and $I_b$, $\omega_B(I_b/I_C,V_b/V_C) = \omega_J(V_b/V_C,I_b/I_C)$. In average, these frequencies are close to those of uncoupled oscillators, $\bar{\omega}_B(V_b), \bar{\omega}_J(I_b)$, the deviations are mostly due to synchronization. We observe the domains corresponding to the fractions $n/m$. They are centred at the curves where $m \bar{\omega}_B(V_b) = n \bar{\omega}_J(I_b)$. The widest domain is that with $n=1,m=1$ and is centered at the diagonal. The domains with higher $n,m$ are increasingly more narrow, as it is also the case in QHE.    
The parameters are chosen such that the resonant frequency $\Omega$ is achieved at $I_b/I_C = V_b/V_C =\sqrt{2}$, were the domains are widest. $R_S=10 \pi \hbar/e^2$ and the oscillator impedance is optimized, $z_0=\sqrt{R_S/G_J}$, so that $Q_m =\sqrt{R_S/G_J}/2= 10$. In accordance with above estimations, the widest synchronization domain spreads at the scale of $\Omega$ itself. The widths of the domains decrease at much higher and much lower frequencies $\bar{\omega}_B$, $\bar{\omega}_J$ owing to decrease of $K(\omega)$. More details and finer steps can be seen in the right pane where the transresistance is plotted along the cut in $V_b-I_b$ plane showing a typical devil's staircase curve. As a side note, the domains are not precisely single-connected, there is a fine structure of small "islands" of the same $n,m$ near each domain. This structure is however too fine to be resolved at the scale of the plots.    

\begin{figure} 
\includegraphics[width=\columnwidth]{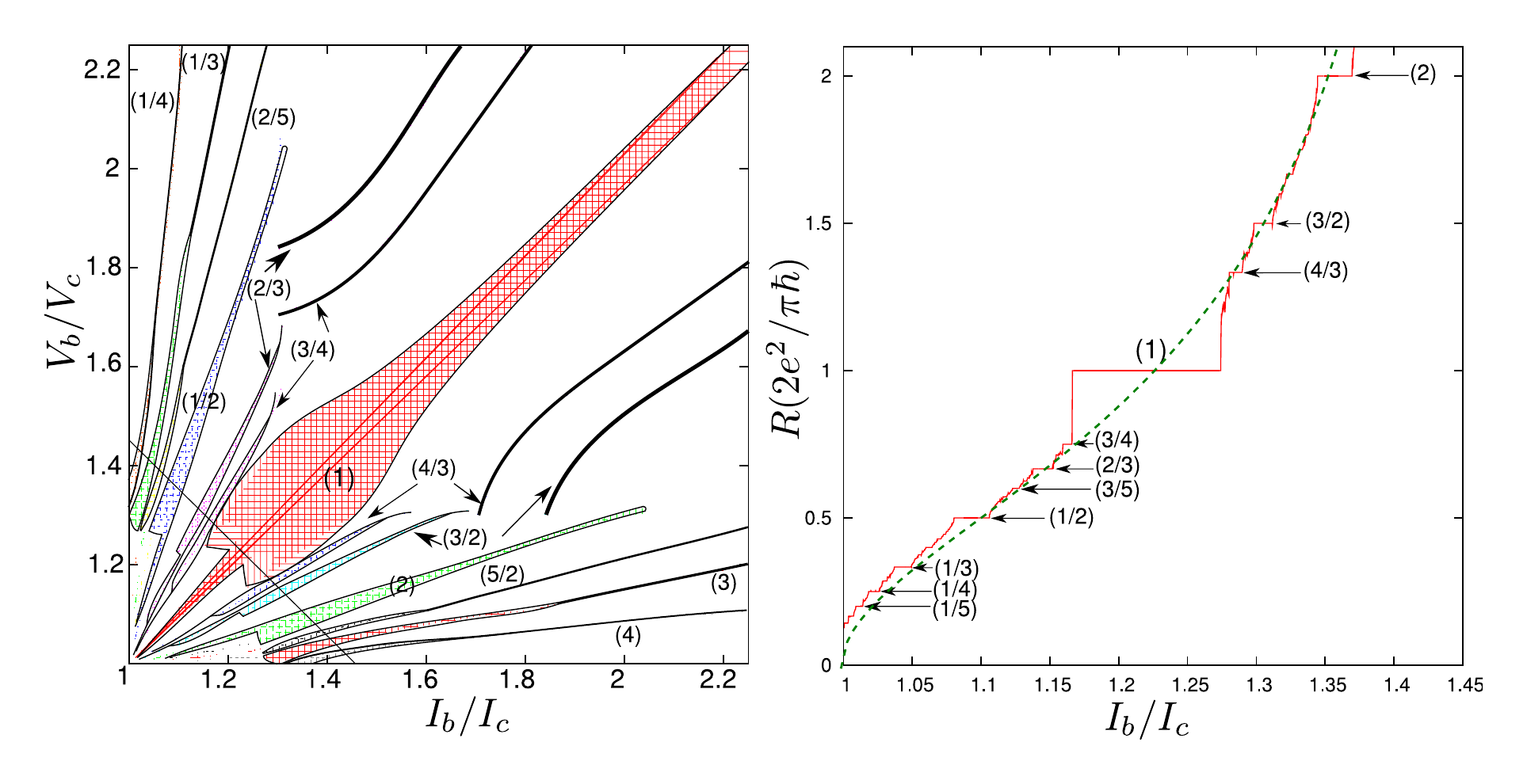}
\caption{Left: Synchronization domains $(n/m)$ in the plane of normalized bias voltage and bias current. Right: Quantized plateaus of transresistance $R = V_{\cal O}/I_{\cal O}$ along the cut given by the line in the left figure. Dashed curve: continuous transresistance as set by uncoupled Bloch and Josephson parts, $R = (\pi\hbar/2e^2) \bar{\omega}_J(I_b)/\bar{\omega}_B(V_b)$.}
\label{Fig3}
\end{figure}

To address the quantum effects, we restrict ourselves
to narrow synchronization domains where a new long time scale $\simeq (\Delta \omega_{B,J})^{-1} \gg (\omega_{B,J})^{-1}$ emerges. At this time scale, one can disregard the dispersion of quantum noise and amplification coefficient and end up with a local-in-time action which is formally equivalent to that of a classical system subject to a white noise. Similar approach has been applied to narrow Shapiro steps \cite{LikharevBook}. The slow variables in our case are the {\it phases} $\theta(t),\Psi(t)$ of Bloch and Josephson oscillations, respectively. With those, the time-dependent current (voltage) is represented as $I_{\cal O}(t) = I_{\cal O} + \tilde{I}_{\cal O}(\bar{\omega}_B t + \theta(t))$ ($V_{\cal O}(t) = V_{\cal O} + \tilde{I}_{\cal O}(\bar{\omega}_J t + \Psi(t))$), $\tilde{I}_{\cal O}$, $\tilde{V}_{\cal O}$ . We derive the effective action in the vicinity of the point in $I_b-V_b$ plane where $n\bar{\omega}_J = m \bar{\omega}_B = \omega$ aiming to describing the $(n,m)$ domain (In formulas for the action, $\hbar=1$ for compactness).
\begin{align}
{\cal S} &= {\cal S}_B + {\cal S}_J + {\cal S}_{cp}; \\
{\cal S}_B&= r \int dt \left(\dot{\theta}\theta_d -i T^*_B \theta_d^2 -(\delta \omega_B)\theta_d \right), \\
{\cal S}_J&= g \int dt \left(\dot{\Psi}\Psi_d -i T^*_J \Psi_d^2 -(\delta \omega_J)\Psi_d \right), \\
{\cal S}_{cp}&=\omega \frac{|K|}{2\pi} \int dt \Big(-A_B \cos(m\theta -n \Psi +\kappa)\cdot\theta_d 
\\&+ A_J \cos(m\theta -n \Psi -\kappa)\cdot\Psi_d \Big).
\end{align}
Here, ${\cal S}_{B,J}$ describe Brownian motion of the phases in the absence of the coupling, $g(r) \gg 1$ being dimensionless differential conductance (resistance), $g\equiv (\hbar/4e^2)(dI_b/dV_{\cal O})$ ($r\equiv (e^2/\pi^2\hbar) dV_b/d I_{\cal O}$), $T^*_{J,B} \simeq \hbar \omega$ being effective noise temperatures that depend on bias current and voltage. ${\cal S}_{sp}$ gives energy  ($\simeq \hbar |K|$)gained  by synchronization, $\kappa \equiv {\rm arg}(K)$.
The coefficients $A_{B,J}$ depend on $I_b,V_b$ as well as on $n,m$. 
We concentrate on the relevant variable $\gamma = m \theta -n \Psi$ to reduce the action to the form 
\begin{align}
{\cal S} = \int dt \left(a(\dot{\gamma}\gamma_d -i T^* \gamma_d^2 - \delta \omega) - E_{cp} \sin \gamma\gamma_d \right). 
\end{align}
Here, the susceptibility $a = gr/(gm^2+rn^2)$, noise temperature $T^* = (T^*_B m^2 g+T^*_J n^2 r)/(gm^2+rn^2)$, the energy barrier $E_{cp} =\hbar \omega |A_B nr K + A_J mg K^*|/(gm^2+rn^2)$,
and $\delta \omega = m \delta \bar{\omega}_B - n \delta \bar{\omega}_J$.
This action is formally equivalent to that of an overdamped particle moving in a trapping washboard potential $U(\gamma) = -E_{cp} \cos\gamma - \gamma \hbar a \delta\omega$ (Fig. 4) and being subject to the thermal noise. If we neglect the noise,
the motion obeys
$a \dot{\gamma} + \partial U(\gamma)/\partial \gamma =0$.
The stationary solutions to this equation where $\gamma$ in trapped in one of the minima correspond to the synchronization of the oscillations. They occur within a strip $|\delta \omega| \Delta \omega \equiv E_{cp}/\hbar a$, in accordance with the estimations made. Beyond the strip, $\gamma$ increases with time corresponding to two unsynchronized frequencies. 

\begin{figure} 
\includegraphics[width=\columnwidth]{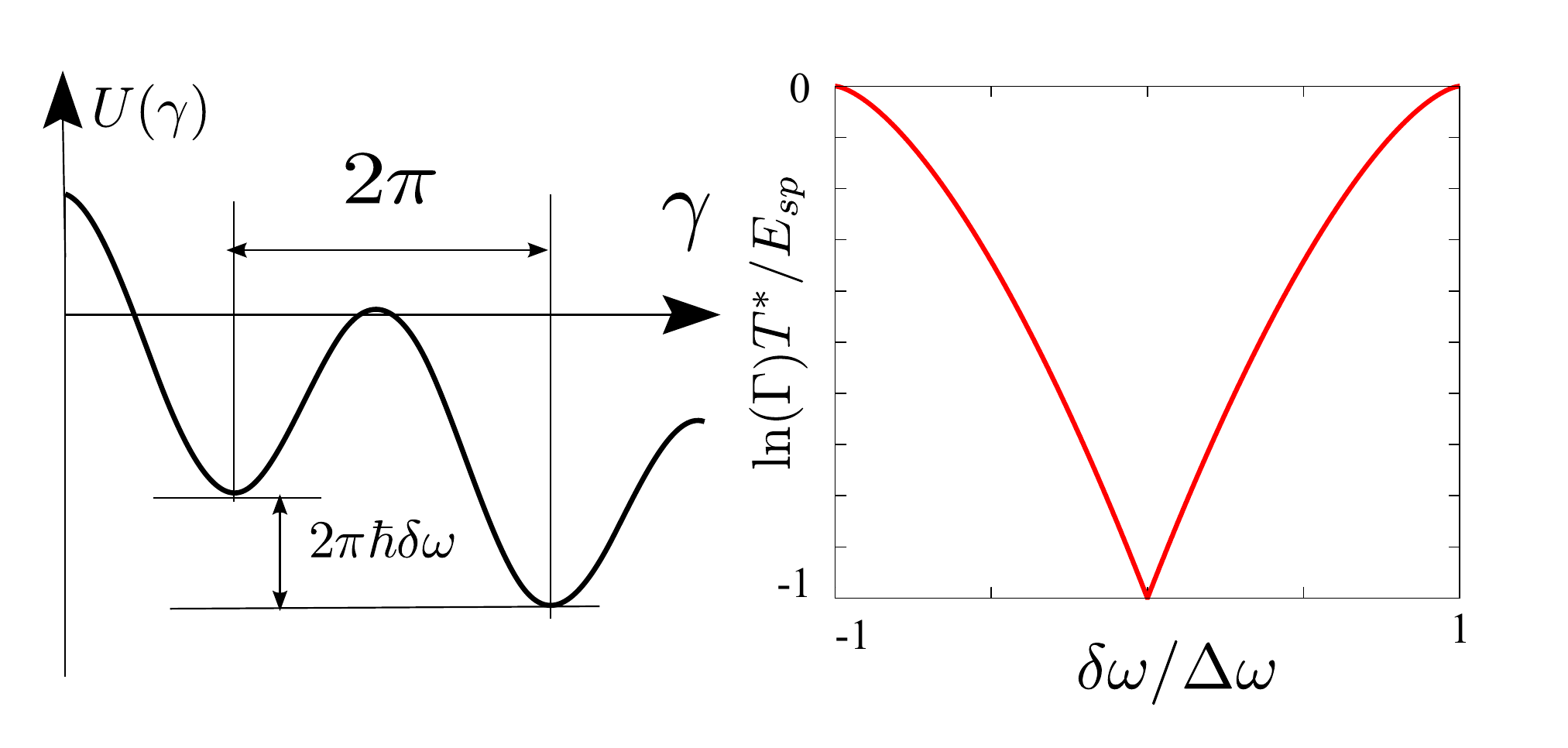}
\caption{Left: Washboard potential for 
the phase difference $\gamma$. The hops over the barriers are synchronization errors. Right: The logarithm of the error rate across the synchronization domain.}
\label{Fig3}
\end{figure}
 
The synchronization errors are thermally-activated hops between the neighboring minima. Their rate determines the accuracy of the resistance quantization. It is clear that this rate is exponentially small,
$\ln \Gamma = -E_{cp}/T^{*}$, in the center of the synchronization domain, this guarantees the high quality of the resistance quantization. The rate increases towards the strip edge owing to the lowering of the barrier in the washboard potential,
$\ln \Gamma = - (E_{cp}/T^*)((1-y^2)+y\arccos(y))$, $y \equiv |\delta \omega|/\Delta \omega$.
The coefficient $E_{cp}/T^* \simeq K$ depends on bias current and voltage as well as on $n,m$.
We provide extensive illustrations of this dependence\cite{supp}. 

In FQHE, the excitation bear fractional charge/flux. The synchronization errors may also be considered as excitations at the background of a synchronization domain. 
One might conjecture that {\it extra} charge/flux induced by a hop over the barrier is fractional: this would be the case if the $2\pi$ change in $\gamma$ is equally split between the phases $\theta,\Psi$. In fact, the situation is more complex since the hop takes a relatively long time $\simeq a\hbar/E_{cp}$ during which the charge and flux (related to the superconducting phase difference $\phi$) may fluctuate. Owing to this, the average extra charge and flux transferred in course of a hop do not exhibit a strict quantization,
\begin{equation}
\frac{\delta q}{2e} = \frac{m g}{g m^2+r n^2};\;\;\frac{\delta{\phi}}{2\pi}= \frac{-n r}{g m^2+r n^2}
\end{equation}
However, in the limit $g \gg r$ the extra charge approaches fractional value $2e/m$, while the extra flux approaches $-1/n$ in the opposite limit.

In conclusion, we have proven the feasibility of synchronization of Bloch and Josephson oscillations in superconducting devices that is manifested as a (fractionally) quantized transresistance like in QH devices. High amplification coefficient is required for the stable synchronization, this is readily achieved by using an $LC$-resonator with high quality factor $Q$. The minimum synchronization error rate is exponential in $Q$.


We acknowledge fruitful discussions with J. E. Mooij, K. K. Likharev, and 
A. V. Zorin. Our research was
supported by the Dutch Science Foundation NWO/FOM.


\onecolumngrid


\begin{center}
\textbf{{\large Supplementary Material for\\
``Quantum synchronization and transresistance quantization in superconducting devices''}}\\
\bigskip
A.~M.~\ Hriscu, Yu.~V.~\ Nazarov\\
\textit{Kavli Institute of NanoScience, Delft University of Technology, 2628 CJ Delft, The Netherlands}
\end{center}

Here we present supplementary material for our article "Quantum synchronization and transresistance quantization in superconducting devices". We give here details of the original Keldysh action that describes a general coupling circuit, specify classical saddle-point equations and details of their analysis. We present the derivation of simple action for narrow strips and complex formulas for the coefficients involved. We illustrate the dependences of the synchronization error rate for various synchronization domains on the parameters of the device proposed.

\section{Details of the general action}
The action for a general circuit where Bloch and Josephson junctions are present consists of a non-linear part
$$
{\cal S}_{nl} = \int dt \left( 2 E_J \sin\phi \sin\frac{\phi_d}{2} + 
2 E_S \sin q \sin\frac{q_d}{2}\right)
$$
the part describing the bias current and voltage
$$
{\cal S}_{bias} = - \int dt \left(\frac{e V_b}{\pi}q_d + \frac{I_b}{2e} \phi_d\right)
$$ 
and the part describing the linear circuit. 
To get  this part of the action , we use general expressions for Keldysh action of a linear system. Given a vector 
 $x_i$ of generalized coordinates, corresponding generalized forces $f_i$, and the susceptibility matrix $\hat\chi(\omega)$
that relates the coordinates and forces,
$$
x_i(\omega) = \sum_{ij} \chi_{ij}(\omega) f_{ij} 
$$
the action can be written as 
($\hat{A} \equiv \hat{\chi}^{-1}$):
$$
{\cal S} = \int \frac{d\omega}{2\pi} \left(x^{d}_{-\omega} \hat{A}_{\omega} x_\omega  - \frac{i}{2} x^{d}_{-\omega} \hat{S}_\omega x^{d}_\omega\right) - f x_d
$$ 
The terms $\propto x x_d$ describe the response as well as dynamics of $x$.
The terms $\propto x^2_d$ describe the quantum and thermal noise. The quantity $\hat{S}$ is a correlator of random forces.  By virtue of fluctuation-dissipation theorem it is expressed in terms of susceptibility as 
$$
\hat{S}_\omega = \frac{i}{2} \left(\hat{A} - \hat{A}^{\dagger}\right){\rm coth} \left(\frac{\omega}{2T}\right).
$$
This matrix is positively defined (eventually, ${\rm Im} A$ is negatively defined at positive frequency)

In our case, the generalized forces are the voltage source connected to the  Bloch terminal and the current source connected to the Josephson terminal, such that the generalized coordinates are related to the charge and flux at the junctions. We use a common convention where positive currents are {\it out} of the terminals. We compute matrix $\hat{U}$ that relates sources and responses, those are current $j$ in Bloch terminal  and voltage $v$ at Josephson terminal. The susceptibility matrix $\hat{\chi} =\hat{U}/(-i\omega)$. The explicit answer for $\hat{U}$ reads 
\begin{eqnarray}
\left[\begin{array}{c} j \cr v \end{array}\right]
= \hat{U} 
\left[\begin{array}{c} V \cr I \end{array}\right];\; \\
\hat{U} = \frac{1}{R_S Z_2 +(R_S+Z_2)(Z_1+G^{-1}_J)}
\left[\begin{array}{cc}
Z_2+Z_1+G^{-1}_J & - Z_2 G^{-1}_J\cr
Z_2 Z_1 & Z_1 (R_S Z_2 +Z_1(R_S+Z_2))
\end{array}
\right].
\end{eqnarray}
Inverting the matrix, we obtain
$$
\hat{A} = -i\omega \left[\begin{array}{cc}
R_S+\frac{Z_2Z_1}{Z_1+Z_2} & \frac{Z_2}{Z_2+Z_1}\cr
-\frac{Z_2}{Z_2+Z_1} & G_J +\frac{1}{Z_2+Z_1}
\end{array}\right] \equiv 
-i\omega \left[\begin{array}{cc}
R_S +\delta R_S & K\cr
-K & G_J +\delta G_J
\end{array}
\right].
$$
Here, like in the main text, we have introduced the amplification coefficient $K \equiv Z_2/(Z_1+Z_2)$ and coupling-induced modifications $\delta R_S = Z_2Z_1/(Z_2+Z_1)$, $\delta G_J = 1/(Z_2+Z_1)$.

The variables $\phi$,$q$ differ by factors from the generalized displacements $\Phi$ and $Q$, $\phi = 2e\Phi$, $Q=(e/\pi) q$.
With this, the part of the action representing the linear circuit reads:
\begin{eqnarray}
{\cal S} &=&{\cal S}_1 +{\cal S}_2 +{\cal S}_3+{\cal S}_4\\
{\cal S}_1&=& \int \frac{d\omega}{2\pi} \left(\phi^d_{-\omega} \frac{\tilde{G}}{4e^2}\left(\dot{\phi}\right)_\omega + q^d_{-\omega} \frac{e^2 \tilde{R}}{\pi^2}\left(\dot{q}\right)_\omega \right)\\
{\cal S}_2&=& \int \frac{d\omega}{2\pi} \frac{K(\omega)}{2\pi}\left(q^d_{-\omega}\left(\dot{\phi}\right)_\omega - \phi^d_{-\omega}\left(\dot{q}\right)_\omega\right) \label{coupling}\\
{\cal S}_3&=& \int \frac{d\omega}{2\pi} \frac{\omega}{2} {\rm coth} \left(\frac{\omega}{2T}\right)\left(\phi^d_{-\omega} \frac{{\rm Re}\tilde{G}}{4e^2}\phi^d_\omega + q^d_{-\omega} \frac{e^2 {\rm Re}\tilde{R}}{\pi^2}q^d_\omega \right)\\
{\cal S}_4&=& \int \frac{d\omega}{2\pi} \frac{\omega}{2} {\rm coth} \left(\frac{\omega}{2T}\right)  i \frac{{\rm Im} K(\omega)}{2\pi} \left(q^d_{-\omega}\phi_{\omega} -q^d_{\omega}\phi_{-\omega}\right)
\end{eqnarray}
with $\tilde{G} = G_J +\delta G_J$, $\tilde{R}=R_S +\delta R_S$.
\section{Classical equations}

The classical equations for the circuit can be obtained either by
varying the action with respect to $\phi_d,q_d$ and setting $\phi_d,q_d=0$ or by applying Kirchoff equations to the circuit. 
The latter will allow us to keep the equations in the differential form 
 
The classical equations for $q$ and $\phi$ are separate in the absence of the coupling,
\begin{align}
I+ I_C &\sin{\phi} + \frac{\hbar}{2e R_J} \frac{d \phi}{d t}=0\\
V+V_C &\sin{ q} + \frac{e}{\pi} R_S \frac{d q}{d t}=0
\end{align}
and are obtained by applying current conservation in Josephson branch 
and summing up the voltage drops along the Bloch branch.
With the coupling elements, these equations include the current in the capacitor $I_{cap}$ and voltage drop $V_L$ over the inductor,
\begin{align}
I+ I_C &\sin{\phi} + \frac{\hbar}{2e R_J} \frac{d \phi}{d t}-I_{cap}=0\\
V+V_C &\sin{ q} + \frac{e}{\pi} R_S \frac{d q}{d t}+V_L=0
\end{align}
that are related to the corresponding current and voltage
by $I_{cap}=C \dot{V}$, $V_L = L \dot{I}_L$.
Two extra equations are obtained by applying current conservation in the node connected to the capacitor and the inductance, and summing up the voltage drops across the Josephson junction, capacitor, and inductance,
\begin{equation}
\frac{e}{\pi} \dot{q} = I_{cap} + I_L ;\; 
\frac{\dot{\phi}}{2e} = -V_{cap} + V_{L}.
\end{equation}

We reduce these 6 equations to 4 equations that express time derivatives of $q,\phi, V_L, I_L$ in terms of these 4 variables. We solve these evolution equations numerically by a rk4 solver. A typical computer run is as follows. We fix the device parameters $E_J,E_S, R_S, G_J, L, C$ for the whole run. We vary changeable parameters $I_b,V_b$ with small steps, either along a line in the $I_b-V_b$ plane or scanning a square in the plane. For each point in the $I_b-V_b$ space we start with an arbitrary initial condition and "wait" (typically, $30-40$ $Q_m/\Omega$) for equilibration of the resulting oscillations. Then we check whether the resulting orbit is a periodic one in the space of $q,\phi, V_L, I_L$ that is, if initially 
at the point $(q,\phi, V_L, I_L)$, it arrives to the point 
$(q+2\pi n,\phi+2\pi m, V_L, I_L)$ with some integer $n,m$ after a time interval. If the periodicity is found, the program outputs the numbers $n$ and $m$ and the point in the $I_b-V_b$ plane is marked as to belong to the synchronization domain $n,m$. Otherwise, the point  is regarded as to belong to the chaotic domain.
In a faster version of the simulation, $I_b$,$V_b$ are continuously and very slowly updated with time, and the periodicity of the orbits is  monitored constantly. 

In this way, we have obtained the plots given in the main text. Those are made with relative resolution  $10^{-4}$ in $I_b, V_b$ and show single-connected synchronization domains at the chaotic background with widths quickly decreasing with increasing ${\rm max}(n,m)$.

If we concentrate at a domain boundary, say, of the $(1,1)$ domain and increase resolution, we find more structure. We are able to see the "paddles" of the chaotic domain within $(1,1)$ domain, the small "islands" of $(1,1)$ synchronization in the chaotic region and even smaller synchronization domains with large $n,m$. The structure exhibit fractal self-similarity upon scaling the resolution and the size of the region scanned. This is what is generally expected from the transition between commensurability and chaos and should not surprise. Since the structure is seen at high relative resolutions only, we expect it , in distinction from wider domains, to vanish at any realistic noise level and therefore did not investigate it in detail. 
 
\section{Derivation of the simplified action}

Our goal is to derive a simplified action 
for slow variables that are the phases $\Psi,\theta$
of the Josephson and Bloch oscillations, correspondingly. One can  draw a similarity with a well-known phenomenon of Shapiro steps in Josephson junction whereby Josephson oscillations are synchronized with an external a.c. current signal.
Our case can be regarded as a sort of spontaneous emergence of such signal whereby the synchronizing signal for Josephson part is produced by the Bloch part, and vice versa. To this end, we start with deriving the action for Shapiro steps in the resistively shunted Josephson junction.   
\subsection{Simplified action for Shapiro steps}
Our starting point is the full action for the resistively shunted Josephson junction,
\begin{equation}
{\cal S} = \int dt \left( 2 E_J \sin\phi \sin\frac{\phi_d}{2} -\frac{I +\tilde{I}(t)}{2e} \phi_d  + \dot{\phi} \phi_d \frac{G}{4e^2}\right) + {\cal S}_{n}
\end{equation}
where the noise term ${\cal S}_{n}$ we write at the moment as
\begin{equation}
{\cal S}_{n} = -i \frac{1}{2} \int  dt dt' \phi_d(t)s(t-t') \phi_d(t')
\end{equation}
without specifying the kernel $s$.

We assume $G \gg e^2/\hbar$, (effective) noise temperature $\ll E_J$, and small a.c. amplitude $\tilde I \ll I_C$. Under these conditions, the realizations of $\phi,\phi_d$ are close to solutions of the stationary saddle-point equation, that is, one without noise, and $\phi_d$ is typically small. The specifics of the situation that these solutions are time-dependent periodic oscillations and therefore are not unique: For each solution $\phi(t)$ there is a phase-shifted solution $\phi(t+\Psi/\omega)$, $\omega$ being the oscillation frequency.
The field $\Psi$ can therefore be regarded as a Goldstone mode. To come to the effective action for $\Psi$, one allows for slow time-dependence of this field, expresses $\phi(t),\phi_d(t)$ in terms of $\Phi(t),\Phi_d(t)$, and substitutes these expressions into the action. This rather straightforward program has some less trivial implementation details outlined below.     

The classical equation without noise is obtained by varying the general action with respect to $\phi_d$. It reads 
\begin{equation}
E_J \sin\phi - \frac{I}{2e} +\frac{G}{4e^2} \dot{\phi} =0
\end{equation}
To solve it, we first make it dimensionless by introducing $I_C = 2e E_J$, dimensionless current $j \equiv I/I_C$, and
dimensionless time $\tau$ such that
\begin{equation}
\frac{d}{d\tau} = \frac{G}{4 e^2 E_J} \frac{d}{dt}.
\end{equation}
With this, the equation reads
\begin{equation}
\label{eq:classic}
\dot{\phi} +\sin\phi -j =0
\end{equation}
To solve it, let us change the variable to $Z \equiv e^{i\phi}$, 
$\dot{Z} = i\dot{\phi}Z$, so the equation becomes
\begin{equation}
\dot{Z} +\frac{1}{2}\left( Z^2 - 1 - 2ijZ\right) =0 \to 
\frac{dZ}{Z^2 - 1 - 2ijZ} = - d\tau/2
\end{equation}
Let us substitute $j = \frac{1}{2}(y+y^{-1})$, thus defining a convenient parameter $y>1$.
With this, we can factorize the denominator,
\begin{equation}
Z^2 - 1 - 2ijZ = \left(Z -i y\right)\left(Z-i y^{-1}\right)
\end{equation}
Integrating both parts we get
\begin{equation}
\ln\left(\frac{Z-iy}{Z-i y^{-1}}\right)\frac{1}{i(y-y^{-1})}  = -\tau/2 +C 
\end{equation}
We choose the constant $C$ in such a way that $Z=1$ at $\tau=0$ and exponentiate to obtain
\begin{equation}
\frac{Z - iy}{Z -i y^{-1}} \frac{1-iy^{-1}}{1-iy} = \exp\left(-i \tau \omega_j \right)\end{equation}
with the dimensionless Josephson frequency $\nu_J = \frac{1}{2}(y-y^{-1}) = \sqrt{j^2-1}$. 
The frequency of correct dimension is then $\omega_J  =\frac{4e^2E_J}{G}\nu_J$.

We can now rewrite the equation for $Z$ as 
\begin{equation}
\frac{Z - iy}{Z -i y^{-1}} = y \exp(-i\tau \omega_J -i\chi) \equiv y X
\end{equation}
introducing a phase factor 
$$\exp(i\chi) \equiv -i \frac{1+iy}{1-iy}$$
to obtain
$$
Z = i \frac{X-y}{yX -1} = iy + i \frac{1-y^2}{y-1/X} = \frac{i}{y} - i \frac{2\omega_J}{yX-1}
$$
That we can expand in $1/X$, that is, in harmonics:
$$
Z = \frac{i}{y} -i 2\omega_J \sum_{n=1} \frac{1}{X^n y^n}
$$
There are only harmonics with negative and zero frequency.

Let us compute the voltage $d\phi/d\tau$. 
It is given by
$$
\frac{\dot{Z}}{i Z} = -i\dot{X} \frac{\partial Z}{\partial X} Z^{-1} = - \nu_J  X\frac{\partial Z}{\partial X} Z^{-1} = - 2 \nu^2_J \frac{yX}{(X-y)(yX-1)} = 2\nu^2_J \frac{y}{y^2+1 - y(X+1/X)}
$$
It is a less expected property of this equation that the inverse voltage has only zero and first harmonics not involving any higher Fourier components. 
\begin{equation}
(d\phi/d\tau)^{-1} = \omega^{-2}_J \left( j - \cos(\omega_J \tau +\chi)\right).
\label{eq:inverse}
\end{equation}

\subsection{Derivation of slow-variable action}
Let us use the solutions found. We  can express the classical field $\phi(t)$ as 
\begin{equation}
\label{eq:subs-for-phi}
\phi(t)=F(t+\Psi(t)/\omega_J).
\end{equation}
$F(t)$ being the solution found above.
It is important to notice that a change of the field $\Psi$ amounts to a slow part of the change of the superconducting phase $\phi$, since the change of $\phi$ per period of Josephson oscillations is $2\pi$. 

To write a similar expression for the quantum field  $\phi_d$ is less trivial. One might conjecture that Eq. \ref{eq:subs-for-phi} is valid for both parts of the Keldysh contour, that is, for $\phi^{\pm}(t)$, and get the expression for $\phi_d$ in this way. This conjecture is however wrong not satisfying the saddle-point equations. To find the true dependence of $\phi_d$ at the time scale of the oscillation period, we need to inspect the "quantum" saddle-point equation  obtained by varying the full action with respect to $\phi$. Assuming small $\phi_d$, the equation in terms of  dimensionless time $\tau$ reads 
$$
\phi_d \cos\phi-\dot{\phi}_d =0
$$
To solve this equation, we notice that 
$$
\frac{d}{d\tau} \ln(\phi_d) = \cos(\phi).
$$
From the other hand, 
\begin{equation}
\dot{\phi} +\sin\phi -j =0 \ \to \ddot{\phi} + \dot{\phi}\cos\phi \to \cos\phi = - \frac{d}{d\tau}\ln\left(\dot{\phi}\right)
\end{equation}
We conclude that $\phi_d = C (d\phi/d\tau)^{-1}$.  We fix the normalization constant $C$ from the condition that the slow variations of $\phi_d$ correspond to variations of $\Psi_d$, so that the average of $\phi_d$ over the oscillation period  $\bar{\phi_d} \equiv \Psi_d$. 
Using the explicit formula for $d\phi/d\tau$, we 
obtain
\begin{equation}
\phi_d(t) = \Psi_d ( 1 - j^{-1} \cos(\omega_Jt+\chi+\Psi)) 
= \Psi_d \left(d\phi/d\tau\right)^{-1} \frac{\nu_J^2}{j}
\end{equation}

Now it is time to substitute $\phi(t),\phi_d$ to the action and average it over the period. In zeroth order, this cancels the action: indeed, this is generally expected from Goldstone fields. We deal with the residual terms one-by-one. The conductance term where we differentiate $\Psi$  when taking the derivative with respect to time, reads as follows 
\begin{equation}
\frac{G}{4e^2} \overline{\dot{\phi} \phi_d} = \frac{G}{4e^2} \dot{\Psi}\Psi_d \frac{\nu_J}{j}\overline{(d\phi/d\tau)(d\phi/d\tau)^{-1}} = \frac{G^*}{4 e^2} \dot{\Psi}\Psi_d
\end{equation}
where $G^{*} = G \nu_J/j$ is, logically enough, is the differential conductance $(dV/dI_b)^{-1}$.
Next we look at the terms that come with the external current. We expand the current in harmonics of frequency $\omega \approx \omega_J$, 
$$
I(t) = I + \sum_{n=1} {\rm Re} \left(\tilde{I}_n e^{in\omega t}\right) 
$$ 
With this, the action should become 
$$
\frac{\Psi_d}{2e} \left( (\Delta I) + \frac{1}{2} \sum_n {\rm Re}\left(I_n (\phi_d)^*_n \right)\right)
$$
where $\Delta I=I -I(\omega)$ is the external current minus the current corresponding to the frequency $\omega$ and $(\phi_d)_n$ are the normalized harmonics of $\phi_d$. It is a rather peculiar property of the model in use that $(\phi_d)_n =0$ except $n=1$, $(\phi_d)_1 = - j^{-1} \exp(i (\Psi+\chi))$. Owing to this, the model predicts Shapiro steps only if the external frequency equals to Josephson frequency.

By no means this is a general situation. Generally, one expects the Shapiro steps at all multiples of Josephson frequency. In fact, this drawback of purely resistive shunting is well-known in the field of Josephson dynamics and arises from the absence of frequency dependence in $G_J$. If a shunting capacitor is added to the model to provide the frequency dependence, such Shapiro steps readily emerge. We see that also in our simulations, where the frequency dependence is provided by the coupling $LC$ circuit. To simulate the  generic situation without a complication of the model in use, we will force 
the presence of actual harmonics in $\phi_d$ at all frequencies
\begin{equation}
(\phi_d)_n = - \alpha_n j^{-n} \exp(i n(\Psi+\chi)).
\end{equation} 
The values of $\alpha_n \simeq 1$ at $n\ne 1$are subject of choice and can be regarded as extra model parameters.


The noise part of the action after the substitution takes the form
\begin{equation}
\frac{\Psi^2_d}{2} \left(S(0) + \frac{1}{2}\sum_n s(n\omega_J)|(\phi_d)_n|^2\right). 
\end{equation}
So we need the noise kernel at multiples of the Josephson frequency.
In accordance with the fluctuation-dissipation theorem, $s(\omega) = \frac{G}{4e^2} \omega {\rm coth}(\omega/2T)$.
We assume quantum limit $\omega_J \gg T$, so that   $S(\omega_J)=\frac{G}{4e^2}|\omega_J|$.

We can rewrite this contribution to the  action as
$
\frac{\Psi_d^2}{2} \frac{G^*}{4e^2} 2T^*
$,
introducing the effective temperature 
of the Josephson oscillations,
$$
T^*_J = \frac{\omega_J}{4j \sqrt{j^2-1}},
$$ 

Collecting all terms, we obtain the effective action in the following form 

\begin{equation}
{\cal S} = \int dt \left( -\frac{\partial U}{\partial \Psi} \Psi_d + \dot{\Psi} \Psi_d \frac{G^*}{4e^2}-i 2 T^*_J \frac{G^*}{4e^2} \frac{\Psi_d^2}{2} \right)
\end{equation}

where the effective potential  $U(\Psi)$ gives the interaction with the a.c. current,
$$
\frac{\partial U}{\partial \Psi} = 
\frac{1}{2e}\left((\Delta I) +\frac{1}{2}\sum_{n=1} {\rm Re}\left(I_n \alpha_n j^{-n}\exp(-in(\Psi +\chi)) \right)\right).
$$

In distinction from the original action, this action is local in time. It can be solved by integration over subsequent time slices. The resulting evolution equation is a Fokker-Planck equation for distribution of $\Psi$, $P(\Psi)$. For us, it is sufficient to note that the action is equivalent to that of an overdamped  classical particle subject to white noise with effective temperature $T^*$. The synchronization errors correspond to thermally activated hops over the barriers in the potential $U(\Phi)$ and the estimation of their rate is given by Boltzmann factor $\exp(-\Delta U/T^*)$.

\subsection{Action for the Bloch part}
We proceed in a similar way to obtain the action for the slow variables $\theta,\theta_d$ of the Bloch oscillations.

The starting point is the action in terms of $Q,Q_d$
\begin{equation}
{\cal S} = \int dt \left( 2 E_S \sin Q \sin\frac{Q_d}{2} -\frac{e}{\pi} V(t) Q_d  + \dot{Q} Q_d \frac{e^2 R_S}{\pi^2}\right) + {\cal S}_{n}
\end{equation}
with the noise term that we write at the moment as
\begin{equation}
{\cal S}_{n} = -i \frac{1}{2} \int  dt dt' Q_d(t)s(t-t') Q_d(t')
\end{equation}

The classical part of $Q$ is expressed as
$$
Q(t)=F(t+\theta(t)/\omega_B),
$$
where $\omega_B$ is now the Bloch frequency. 
The slow field $\theta$ that adds to the phase of the oscillations. The quantum part is obtained as  

$$
Q_d(\tau) = \theta_d ( 1 - \frac{1}{v} \cos(\omega_B\tau+\chi+\theta_d)) 
= \theta_d \left(dQ/d\tau\right)^{-1} \frac{\nu_B^2}{v}
$$
where we introduce dimensionless bias voltage $v=V_b/V_C$ and dimensionless Bloch frequency $\nu_B = \sqrt{v^2 -1}$.

Again we need to substitute the above expressions $Q(t),Q_d$ to the action and average it over the period. We proceed as in the case of Josephson junction. Let us explicitly concentrate on the dissipative part.
According to fluctuation-dissipation theorem, $S(\omega) = \frac{e^2 R_S}{\pi^2} \omega {\rm coth}(\omega/2T)$.
Assuming $\omega_B \gg T$, we can rewrite the action as
\begin{equation}
\frac{\theta^2_d}{2} S^{*} = \frac{\theta^2_d}{2} 2 T^*\frac{e^2 R_S^*}{\pi^2} 
\end{equation}
introducing effective temperature 
$$
 T^*_B = \frac{\omega_B}{4v \sqrt{v^2-1} }.
$$ 
With this, the effective action reads

\begin{equation}
{\cal S} = \int dt \left( -\frac{\partial U}{\partial \theta} \theta_d + \dot{\theta} \theta_d \frac{e^2 R_S^*}{\pi^2}-i 2 T^*\frac{e^2 R_S^*}{\pi^2} \frac{\theta_d^2}{2} \right)
\end{equation}

where the effective potential  $U(\theta)$ describes the synchronization with the a.c. voltage signal,
$$
\frac{\partial U}{\partial \theta} = 
\frac{e}{\pi}\left((\Delta V) +\frac{1}{2}\sum_{n=1} {\rm Re}\left(V_n \beta_n v^{-n}\exp(-in(\theta +\chi)) \right)\right).
$$
Here, we add $\beta_n$ that are the harmonics of $Q_d$ at $n\ne1$ describe the synchronization at multiples of Bloch frequency. The differential conductance and effective temperature read
$$
R_S^*=R_S\frac{\nu_B}{v}; \quad T^*= \frac{\omega_B}{4v \sqrt{v^2-1} }.
$$

\subsection{Coupling the parts}
Now we can combine the previously derived slow-variable actions for Bloch and Josephson parts. The only remaining work to substitute the harmonics of $Q_d,\phi_d, \dot{Q}, \dot{\phi}$ into the coupling part of the full action (Eq. \ref{coupling}).  
We concentrate on the vicinity of a curve in the $I_b-V_b$ plane where $n \omega_J(I_b) = m \omega_B(V_b) = \omega$. In the vicinity, we can disregard the details of frequency dependence of $K(\omega)$ replacing it with a (big) complex number $K$.  In principle, the synchronization can be achieved at all multiples of the frequency $\omega$. This would however involve $K(\omega)$ at multiples of $\omega$. Since we assume that $K$ is big only in the vicinity of the resonant frequency, we can safely disregard these terms.

The harmonics we need are 
\begin{align}\label{eq:phi}
\phi_d(t)= -\frac{\alpha_n}{2 j^n}\Psi_d \left[e^{in(\omega_J t+\Psi+\chi_J)}+e^{-in(\omega_J t+\Psi+\chi_J)} \right], \\
\label{eq:Q}
Q_d(t)= -\frac{\beta_m}{2 v^m}\theta_d \left[e^{im(\omega_B t+\theta+\chi_B)}+e^{-im(\omega_B t+\theta+\chi_B)} \right].\\
\label{eq:vol}
(\dot{\phi})=\frac{\omega_J^d} {(j+\nu_J)^n}\left(e^{in(\omega_Jt+\Psi+\chi_J)}+ e^{-in(\omega_Jt+\Psi+\chi_J)}\right),
\\
\label{eq:cur}
(\dot{Q})=\frac{\omega_B^d} {(v+\nu_B)^n}\left(e^{im(\omega_Bt+\theta+\chi_B)}+ e^{-im(\omega_Bt+\theta+\chi_B)}\right).
\end{align}

With this, we arrive at 

\begin{equation}
{\cal S}_{cp} =\omega \frac{|K|}{2\pi} \int dt (-A_B \cos(m\theta -n \Psi +\kappa)\theta_d 
+ A_J \cos(m\theta -n \Psi -\kappa)\Psi_d)
\end{equation}
with 
\begin{equation}
A_J = \frac{\alpha_n}{m j^{n} (v+\nu_B)^m};\;
A_B = \frac{\beta_m}{n v^{m}(j+\nu_J)^n}. 
\end{equation}
The phases $\chi_{J,B}$ are irrelevant and can be canceled by corresponding shifts of $\Psi,\theta$. This does not apply to the phase $\kappa$ of the amplification coefficient. 

When deriving the action, we have disregarded ${\cal S}_4$ 
that describes the correlation of the noises induced by coupling. This term is small in comparison with the noise produced by $R_S$, $G_J$
provided  $\delta R_S,\delta G_J \ll R_S,G_J$: this is what we assume.

\subsection{Reduction to the single variable}
The resulting action 
\begin{align}
{\cal S} &= {\cal S}_B + {\cal S}_J + {\cal S}_{cp}; \\
{\cal S}_B&= r \int dt \left(\dot{\theta}\theta_d -i T^*_B \theta_d^2 -(\delta \omega_B)\theta_d \right), \\
{\cal S}_B&= g \int dt \left(\dot{\Psi}\Psi_d -i T^*_J \Psi_d^2 -(\delta \omega_J)\Psi_d \right), \\
{\cal S}_{cp}&=\omega \frac{|K|}{2\pi} \int dt (-A_B \cos(m\theta -n \Psi +\kappa)\theta_d 
+ A_J& \cos(m\theta -n \Psi -\kappa)\Psi_d).
\end{align}
depends on two pairs of variables while the only variable that enters the action in a non-linear fashion is $\gamma \equiv  m \theta -n \Psi$. 
To reduce the action to that of this relevant variable, we 
substitute 
\begin{eqnarray}
\Psi=-\frac{\gamma}{2 n}-\frac{\sigma}{n}; \;\; \theta=+\frac{\gamma}{2 m}-\frac{\sigma}{m}; \\
\Psi_d=-\frac{\gamma_d}{2 n}-\frac{\sigma_d}{n}; \;\; \theta=\frac{\gamma_d}{2 m}-\frac{\sigma_d}{m};
\end{eqnarray}
introducing the fields $\gamma,\gamma_d$, $\sigma,\sigma_d$. The resulting action is quadratic in $\sigma,\sigma_d$ so that this field can be integrated out. Since the integral is Gaussian, the integration is equivalent to finding an optimum in $\sigma,\sigma_d$ and substituting the optimal values $\sigma,\sigma_d$ found back into the action. 

In this way, we arrive at
\begin{align}
{\cal S} = \int dt \left(a(\dot{\gamma}\gamma_d -i T^* \gamma_d^2 - \delta \omega) - E_{cp} \sin \gamma\gamma_d \right). 
\end{align}
Here, the susceptibility $a = gr/(gm^2+rn^2)$, noise temperature $T^* = (T^*_B m^2 g+T^*_J n^2 r)/(gm^2+rn^2)$, the energy barrier $E_{cp} =\hbar \omega |A_B nr K + A_J mg K^*|/(gm^2+rn^2)$,
and $\delta \omega = m \delta \bar{\omega}_B - n \delta \bar{\omega}_J$.
Expressing changes of the fields $\Psi,\theta$ in terms of $\gamma$ we arrive at the relations
\begin{equation}
\frac{\delta q}{2e} = \frac{m g}{g m^2+r n^2};\;\;\frac{\delta{\phi}}{2\pi}= \frac{-n r}{g m^2+r n^2}
\end{equation}
that give the change of the charge and flux passed in the output circuits upon hopping over the potential barrier.

\section{Numerical illustrations}

In the main text, we have argued that high quality factors $Q$, or, equivalently, large amplification coefficients $K$ are required for exponentially good synchronization, 
$\ln \Gamma \propto |K| < Q$.
The effective quality factor is limited by $\sqrt{R_S G_J}/2$ and can therefore be sufficiently large under assumption of well-defined oscillations
$R_S \gg \hbar/e^2$, $G_J \gg e^2/\hbar$.
It is important to find numerical estimates on proportionality coefficient in the above relation. Apparently, the coefficient depends very much on the parameters of the model and tend to be very small for exotic synchronization plateaus with high $n,m$. Here, we present these numerical estimates.

We concentrate on evaluating
a dimensionless coefficient ${\cal A}$ defined by 
$$
\Gamma \simeq \exp\left(-{\cal A} \frac{|K|}{2\pi}\right),
$$ 
$\Gamma$ estimating the synchronization error rate in the middle of a synchronization domain. Thus defined ${\cal A}$ does not depend on the characteristics of $LC$ resonator. 
It follows from the formulas given that 
\begin{equation}
{\cal A} = \frac{E_{cp}}{T^*}\frac{2\pi}{|K|} = \frac{\sqrt{A^2_B n^2 r^2+A^2_B n^2 r^2+ 2 A_JA_B n m g r \cos(2\kappa)}}{\frac{T^*_B}{\hbar\omega} m^2 g + \frac{T^*_J}{\hbar\omega} n^2 r}
\end{equation}
$\chi$ being the phase of the amplification coefficient. All coefficients $A_{B,J},T^*_{B,J}, r,g$
depend on $n,m$ and dimensionless 
voltage and current biases $j=I_b/I_c$,
$v=V_b/V_C$,
\begin{eqnarray}
A_J = \frac{\alpha_n}{m j^{n} (v+\sqrt{v^2-1})^m};\;
A_B = \frac{\beta_m}{n v^{m}(j+\sqrt{j^2-1})^n};\\
r = \bar{r} \frac{\sqrt{v^2-1}}{v};\; g = \bar{g} \frac{\sqrt{j^2-1}}{j};\; {\rm where}\;\bar{r} \equiv(e^2/\pi^2\hbar) R_S;\bar{g} =(\hbar/4e^2)G_J \\
\frac{T^*_B}{\hbar\omega} = \frac{1}{4 m v\sqrt{v^2-1}};\;\frac{T^*_J}{\hbar\omega}= \frac{1}{4 n j\sqrt{j^2-1}}.
\end{eqnarray}

We introduce two dimensionless parameters that characterize asymmetry of the Bloch and Josephson parts,
$$
X = \bar{r}/\bar{g}, Y = E_J/E_S.
$$ 
The symmetry is achieved at $X=Y=1$.

The biases are related
by synchronization condition 
$n \omega_J= m \omega_B$
that in dimensionless notations reads
$$
\sqrt{v^2-1} = XY\frac{n}{m}\sqrt{j^2-1}.
$$

\subsection{Main synchronization domain $n=m=1$}
Let us concentrate first on the main synchronization domain $n=m=1$. We plot in Fig. 1 the coefficient ${\cal A}$ versus the dimensionless bias current $j$.

\begin{figure} 
\includegraphics[width=\columnwidth]{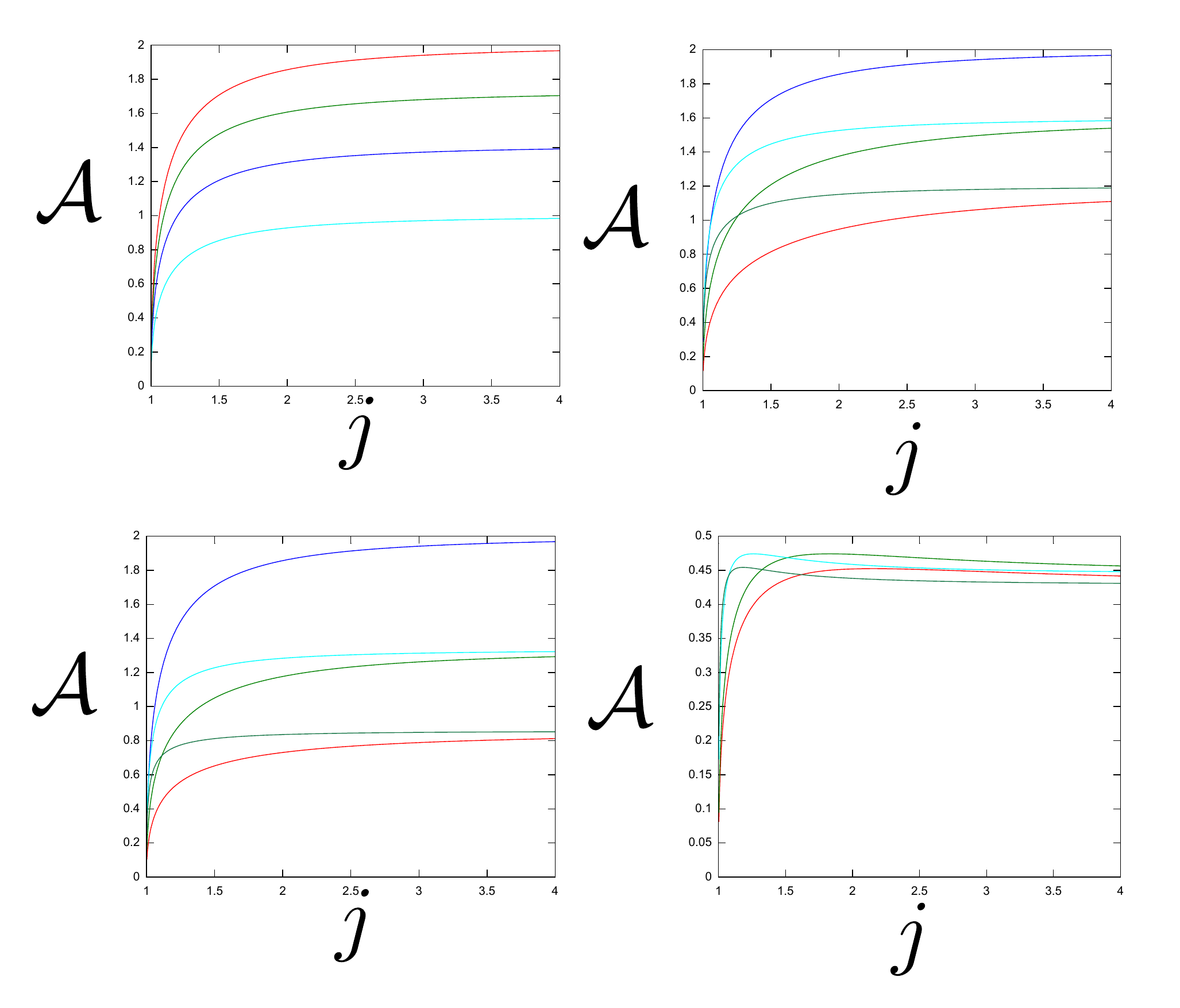}
\caption{The coefficient ${\cal A}$ for main synchronization domain $n=m=1$ versus the dimensionless current bias $j$. Upper left: symmetric setup $X=Y=1$. The phase of the amplification coefficient takes values $\cos(2\kappa) = 1,0.5,0,-0.5,-1.0$ from the upper to the lower curve. Upper right: real $K$, $X=1$ and asymmetry parameter $Y$ takes values $1,2,0.5,3,0.33$ from upper to lower curve at large $j$. Lower left: real $K$, $Y=1$ and asymmetry parameter $X$ takes values $1,2,0.5,3,0.33$ from upper to lower curve at large $j$. Lower right: Purely imaginary $K$, $Y=1$, $X$ takes values $0.5,2,0.33,3$ from upper to lower curve at large bias.}
\label{Sfig1}
\end{figure}

Upper left plot shows the coefficient for different $\cos(2\kappa) = 1,0.5,0,-0.5,-1.0$
under symmetry conditions $X=Y=1$.
At any $j$, the coefficient reaches the maximum  for purely real $K$($\cos(2\kappa)=1$) and vanishes
for purely imaginary $K$, this being an artifact of the symmetry of the setup. 
For all $\cos(2\kappa)$ ${\cal A}$ is zero at the threshold $j=1$ and monotonically increases saturating at big $j$. This is explained by a rather peculiar dependence of $T^{*}/\hbar \omega$ on the bias. This quantity diverges near the threshold and quickly drops with increasing the bias. This drop compensates decrease of $E_{cp}$ at big biases. We note that the strip width that does not depend on $T^{*}$ becomes smaller upon increasing $j$. 

Upper left and lower right plots show the influence of asymmetry on the coefficient, for different $Y$ and $X$, respectively. In all cases, asymmetry decreases the coefficient, and ${\cal A}$ vanishes in the limit of either big or small $X$ as well as in the limit of either small or big $Y$. The reason for this is that at big asymmetries the synchronization condition implies that one of the parts is biased close to the threshold where the effective temperature is high. 

Asymmetry improves ${\cal A}$ in the case of purely imaginary impedance when it vanishes in the symmetric case. This is illustrated in the lower right plot where it is shown that ${\cal A}$ attains a relatively big value $\simeq 0.5$ for not so big asymmetries.

The maximum ${\cal A} =2$ is achieved in the case of real impedance, symmetric conditions and large bias.

\subsection{Domains $n=1$, $m\ne 1$ or $m=1$, $m\ne 1$ }
Let us consider synchronization
domains with higher $n,m$. Among those, the case with either $n=1$ or $m=1$ is rather special. In this case, we also expect steps in our high-damping limit model without adjusting $\alpha,\beta$, so we set $\alpha_p=\beta_p=0$ except $p=1$. In this case, either $A_J$ or $A_B$ equals zero, and the coefficient ${\cal A}$ does not depend on the phase of the amplification coefficient. 
\begin{figure} 
\includegraphics[width=\columnwidth]{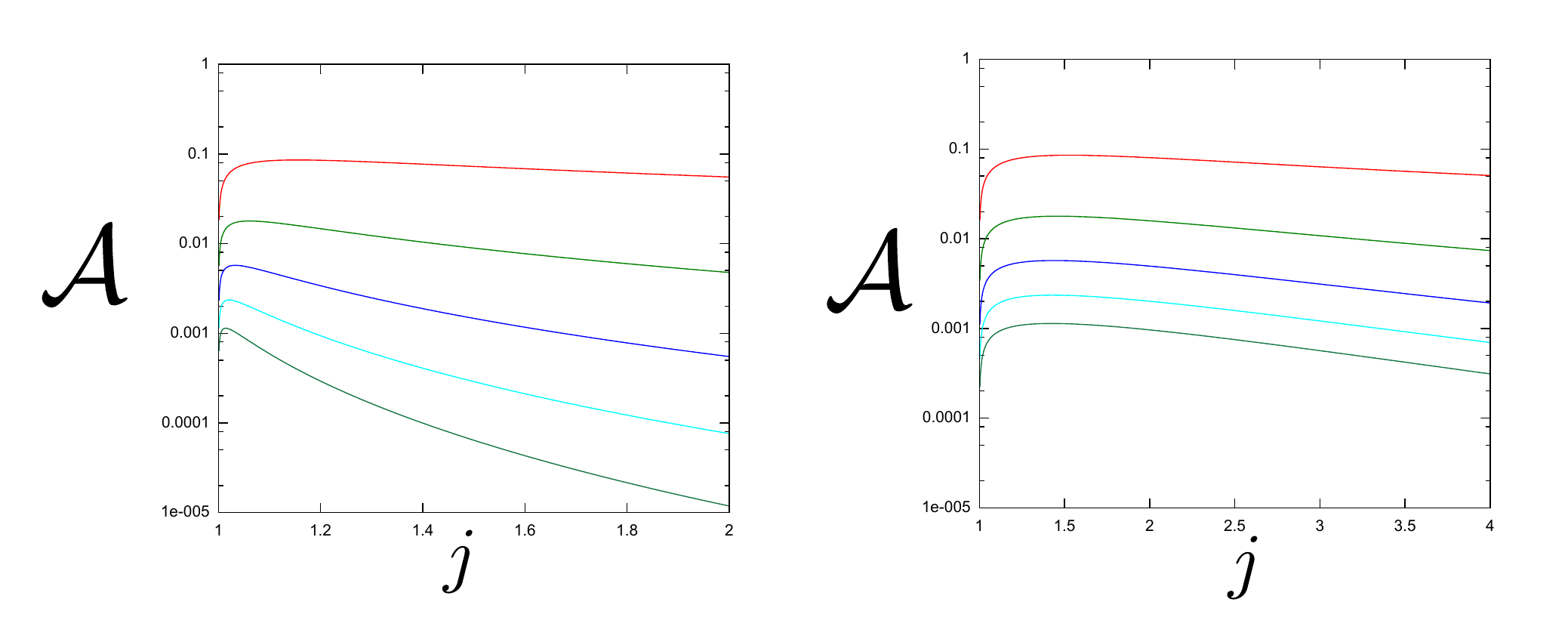}
\caption{The coefficient ${\cal A}$ for  synchronization domains $n=1,m\ne 1$ and  $m=1,n\ne 1$ versus the dimensionless current bias $j$. The coefficient does not depend on the phase of the amplification coefficient, and we assume symmetric case $X=Y=1$. Left: $m=1$, $n$ changes from $2$ to $6$ from upper to lower curve. Right: $n=1$, $m$ changes from $2$ to $6$ from upper to lower curve.}
\label{Sfig2}
\end{figure}
We plot in Fig. 2 ${\cal A}$ versus $j$ for a variety of $n,m$ (note log scale of the vertical axis). We consider only symmetric setup $X=Y=1$. The behavior of the coefficient in $n=1,m$ and $m=1,n$ domains is rather similar so for concreteness we concentrate on $m=1,n$. As in the previous case, the coefficient vanishes at the threshold. Unlike the previous case, it drops upon increasing the bias current. The reason for this is that $E_{cp}$ is proportional to the $n$-th harmonics of the oscillation that scales as $j^{n}$ at large bias. The coefficient ${\cal A}$ therefore reaches the maximum at some $j_n$.
With increasing $n$ $j_n$ approaches the threshold, while the value of the maximum quickly drops. Roughly, it is decreased by a factor of $3$ if $n$ is increased by $1$.

The maximum ${\cal A}$ found is $\approx 0.1$ for $n=1,m=2$ and $m=1,n=2$ domains.

\subsection{Domains $n\ne 1$,$m\ne 1$}

For completeness, we illustrate ${\cal A}$ for 
two general domains: $n/m = 2/5$ and $n/m=3/4$. We set $\alpha=\beta=1$ for all harmonics. 
\begin{figure} 
\includegraphics[width=\columnwidth]{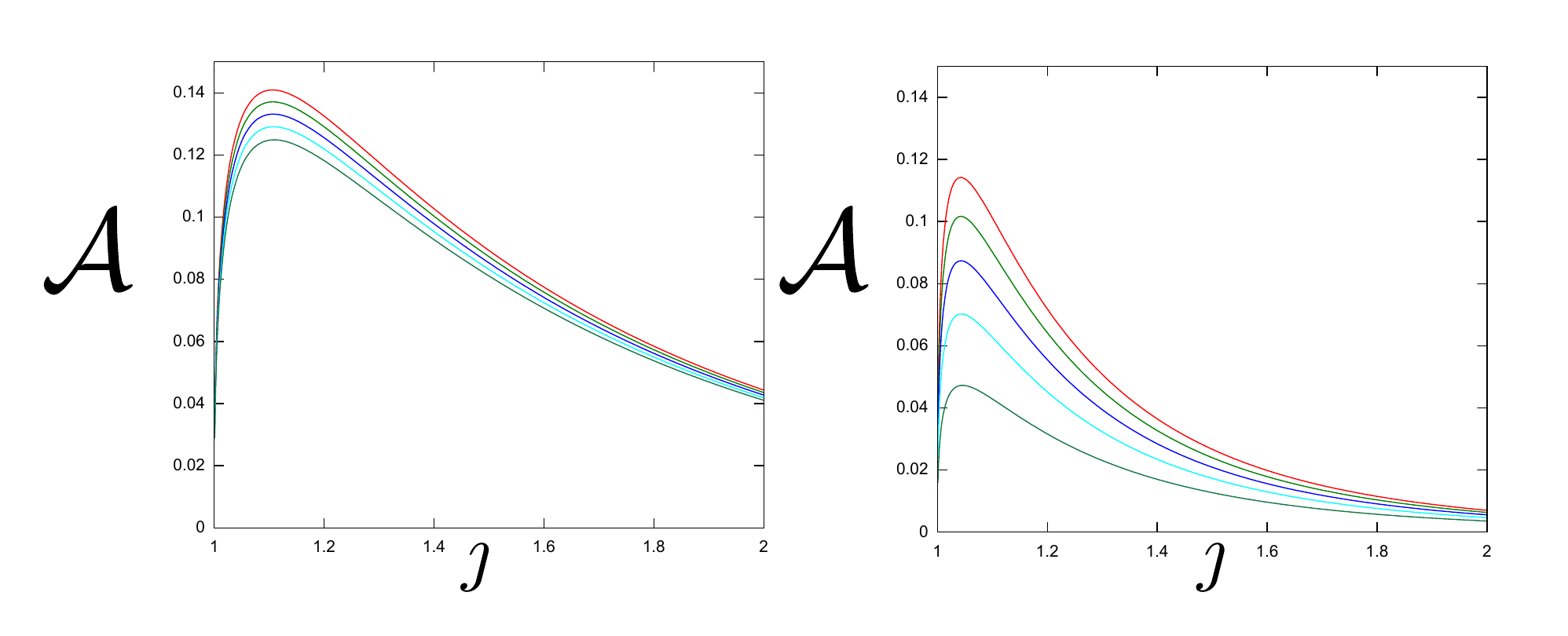}
\caption{The coefficient ${\cal A}$ for  synchronization domains $n/m=2/5$ (left) and  $n/m=3/4$ (right) versus the dimensionless current bias $j$. We assume symmetric case $X=Y=1$. $\cos(2\kappa)$ takes values $1,0.5,0,-0.5,-1$ from upper to lower curve.}
\label{Sfig3}
\end{figure}

The coefficient ${\cal A}$ approaches zero at the threshold and peaks rather close to it dropping down at larger bias currents. The dependence on the phase of the amplification coefficient is bigger for $n/m=3/4$ step since this ratio is close to $1$ where in symmetric case the cancellation of ${\cal A}$ takes place at purely imaginary amplification coefficient. In both domains, maximum value of ${\cal A}$ $\approx 0.15$.

\end{document}